\input{aipcheck.tex}

\def\selectedoptions{final}

\documentclass[\selectedoptions]{aipproc}

\usepackage{psfig}

\def\selectedlayoutstyle{6x9}
\layoutstyle\selectedlayoutstyle

\newcommand\doingARLO[2][]{%
  \ifx\mmref\undefined #1\else #2\fi
}
  
\newcommand\chandra{{\em Chandra}}
\newcommand\asca{{\em ASCA}}
\newcommand\xmm{{\em XMM-Newton}}
\newcommand\sax{{\em BeppoSAX}}

\newcommand\hess{{\em H.E.S.S}}
\newcommand\intg{{\em INTEGRAL}}
\newcommand\cang{{\em CANGAROO}}
 
\newcommand\casa{{Cas\,A}}

\newcommand\snts{{SN\,1006}}
\newcommand\rcwes{{RCW\,86}}
\newcommand\rxjSNR{{RX~J1713.7-3946}}

\newcommand\net{$n_{\rm e}t$}
\newcommand\netunit{cm$^{-3}$s}
\newcommand\tiff{{$^{44}$Ti}}

\newcommand\gray{{$\gamma-$ray}}
                                                                                \newcommand\arcmin{\mbox{$^\prime$}}%
\newcommand\arcsec{\mbox{$^{\prime\prime}$}}%

\def\nat{{Nat\,}}
\def\adspr{{Adv. of Space Research\,}}
\def\apj{{ApJ\,}}
\def\apjl{{ApJL\,}}

\def\aap{{A\&A\,}}
\def\aaps{{A\&AS\,}}

\def\pasj{{PASJ\,}}
\def\aj{{AJ\,}}
\def\mnras{{MNRAS\,}}
                
\newcommand{\kms}{{km\,s$^{-1}$}}

\begin{document}

\title 
      [Non-thermal X-ray Emission From SNRs]
      {Non-thermal X-ray Emission from\\ Supernova Remnants}


\author{Jacco Vink}{
  address={SRON National Institute for Space Research, Sorbonnelaan 2, 3584CA, 
           Utrecht, The Netherlands},
  email={j.vink@sron.nl},
}

\copyrightyear  {2004}

\begin{abstract}
Recent studies of narrow, X-ray synchrotron radiating filaments surrounding
young supernova remnants indicate 
that magnetic fields strengths are relatively high, $B\sim 0.1$~mG,
or even higher, and that diffusion is close to the Bohm limit. 
I illustrate this using \casa\ as an example.
Also older remnants such as \rcwes\ appear
to emit X-ray synchrotron radiation, but the emission is more 
diffuse, and not always confined to a region close to the shock front.  
I argue that for \rcwes\
the magnetic field is likely to be low ($B\approx 17~\mu$G), 
and at the location
where the shell emits X-ray synchrotron radiation the shock velocity is
much higher than the average shock velocity of $\sim 600$~\kms.
\end{abstract}

\date{\today}

\maketitle

\section{Introduction}
Supernova remnants (SNRs) are believed to be the dominant
source of Galactic cosmic rays,
at least for energies up to $3\times 10^{15}$~eV, corresponding
to the ``knee'' in the cosmic ray spectrum, and possibly even up to
$10^{19}$~eV, the ``ankle'' \citep{nagano00}. However, until
a decade ago the best direct evidence for
cosmic ray acceleration by SNR shocks consisted of radio synchrotron
radiation from shell-type\footnote{I single out
shell-type SNRs, as the radiation from 
those remnants are not dominated by synchrotron
radiation from particles accelerated by the pulsar magnetosphere.}
SNRs, providing evidence for the presence
of accelerated electrons with energies in the GeV range, 
far short of the ``knee'' energy, and not providing any evidence for
the acceleration of nuclei, 
which dominate the cosmic ray spectrum.

Recent progress has been made due to major advances
in two fields of high energy astrophysics. 
One is the development of atmospheric Cerenkov detectors,
which are now capable of imaging  TeV \gray s  arcminute resolution,
as shown by the new, exciting image of the
SNR \rxjSNR\ obtained by the High Energy Stereoscopic System (\hess).
The other progress is in the field of 
X-ray astronomy, and in particular the emergence
of imaging spectroscopy with CCD detectors. The first satellite mission 
employing these detectors was \asca, but with \chandra\ imaging 
spectroscopy is now possible at the sub-arcsecond level.
Imaging spectroscopy with \asca\ facilitated  
the first discovery of X-ray synchrotron radiation from
a shell-type SNR, \snts, as it 
allowed for the spatial separation of X-ray line emission
from pure continuum emission \citep{koyama95}.

The detection of X-ray synchrotron radiation from SNRs
provides direct evidence for
electron acceleration up to energies of  10-100 TeV.
Although it does not per se proof the acceleration of cosmic ray nuclei  
by SNRs, and the electron energies fall short of the ``knee'' energy,
it does proof that SNRs are capable of efficient particle acceleration.
Moreover, the combination of X-ray synchrotron and 
TeV \gray\ radiation puts constraints on the relative contributions
of several \gray\ radiation mechanisms, i.e.
$\pi^0$-decay, bremsstrahlung 
or inverse Compton emission, as was done for 
\rxjSNR\ (G347.3-0.5) or \casa\ \citep[][]{lazendic04,aharonian01,vink03a}. 
This is of considerable interest as an unambiguous detection
of $\pi^0$-decay would be direct evidence for the acceleration of
nuclei.

Finally,
the morphology of X-ray synchrotron radiating filaments
in young supernova remnants
gives new insights in the acceleration properties of SNR shock fronts,
e.g. in relation to the idea that cosmic ray streaming leads to significant
magnetic field amplification near high Mach number shocks
due to non-linear growth of plasma waves
\citep{bell01}.
I will discuss this topic using the youngest known Galactic SNR, \casa, as an
example.

Although, it may not be surprising
that young SNRs can efficiently accelerate electrons  up to 10~TeV, 
it is surprising that some apparently
older remnants also
appear to be X-ray synchrotron sources. Moreover, the non-thermal X-ray
emission from these remnants is more diffuse and not confined to 
narrow regions close to the shock front.
Examples are \rcwes\ \citep{vink97,bamba00,rho02}, 
\rxjSNR\footnote{
The age and shock velocity of \rxjSNR\ is highly uncertain.}
\citep{lazendic04,cassam04b}, and even the superbubble 30 Dor C 
\citep{bamba04,smith04}.
As these remnants have slower shock velocities,
which implies a lower  acceleration rate, one would expect that synchrotron
losses would prevent synchrotron radiation to occur in the X-ray band.
In this context I will present some preliminary results of 
\xmm\ and \chandra\ observations of \rcwes.

\section{Cosmic ray acceleration at \casa's fast shocks}
\casa\ is the brightest radio source in the sky and has been for a long time
the archetypal young shell-type SNR. The supernova event may have been dim,
as there are no historical records of a bright supernova 
near the approximate year of explosion  AD 1671
\citep{stephenson02,thorstensen01}.
The estimate for the 
explosion date is based on the kinematics of fast moving optical
knots. However, a more reliable estimate of the current shock velocity
comes from X-ray expansion measurements, which give a shock velocity of
$\sim 5000$~\kms\ \citep{vink98a,delaney03}. The distance to \casa\
is 3.4~kpc \cite{reed95}.

The first evidence for non-thermal X-ray emission was the detection
of  hard X-ray emission above 10~keV by several instruments 
\citep{the96,favata97,allen97}, but it is still debated whether
the non-thermal, hard  X-ray emission is indeed synchrotron radiation 
\citep{bleeker01,vink03a} (see below).

\begin{figure}
\hbox{
  \vbox{
    \psfig{figure=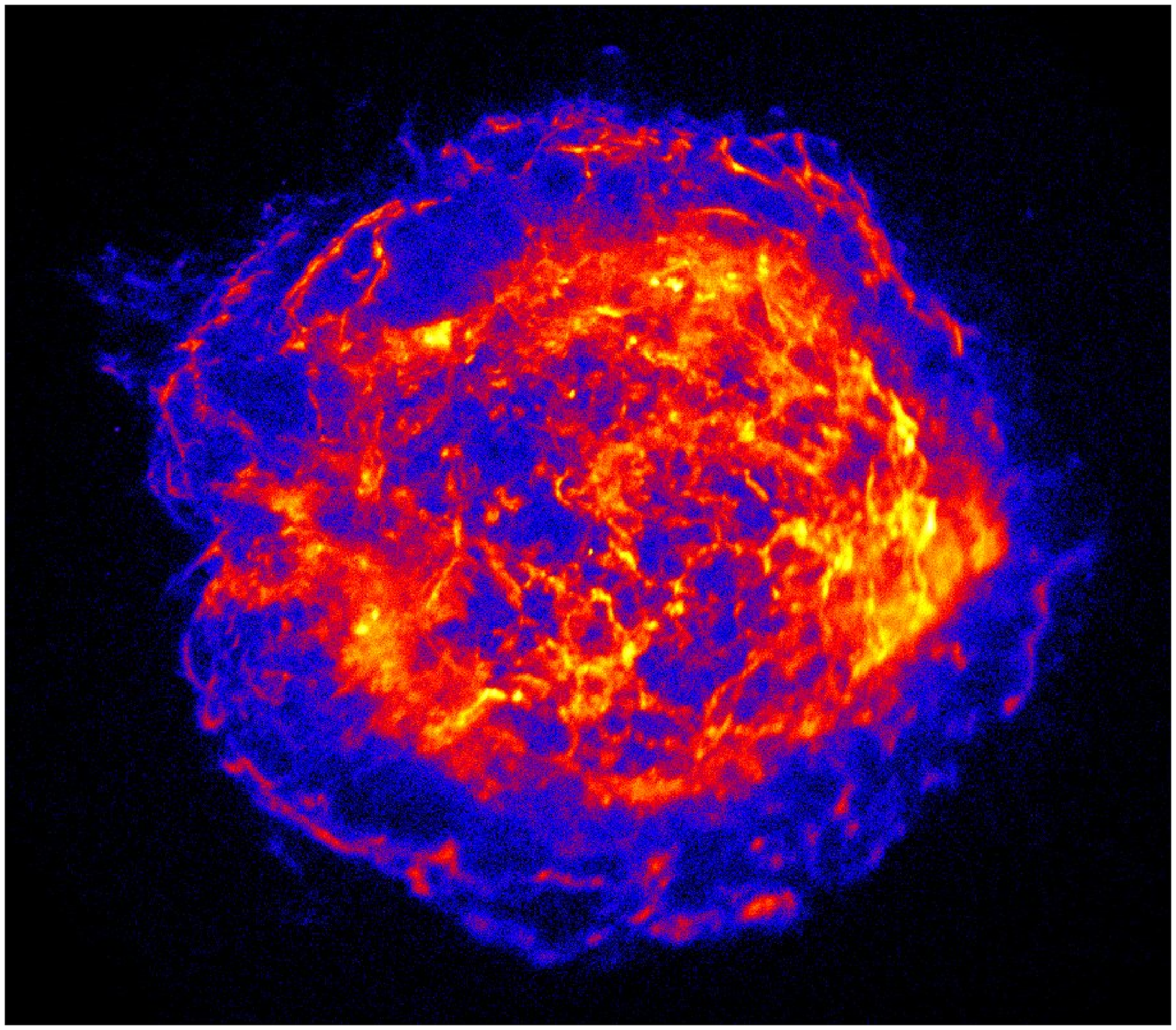,width=0.5\textwidth}
    \vskip 8mm
  }
  \psfig{figure=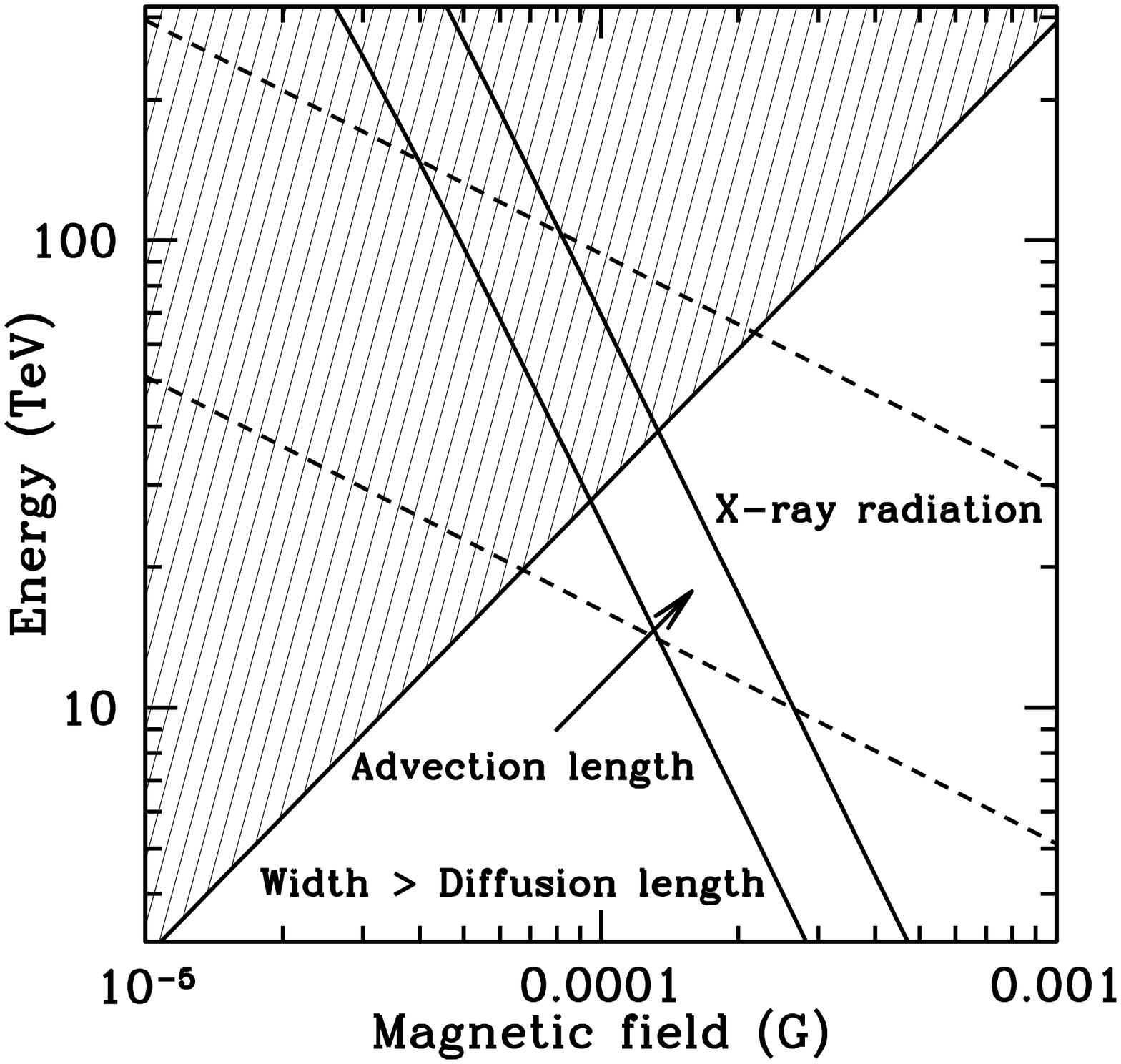,width=0.5\textwidth}}
  \caption{A recent \chandra\ image \citep{hwang04}
in the 4-6 keV continuum band, intensities have been logarithmically
scaled  (left).
Note the thin filaments marking the border of the remnant
(NB the point spread function is not 
uniform). The remnant has a radius of about 2.5\arcmin.
Right: Determination of the maximum electron 
energy versus magnetic field strength 
for the region just downstream of \casa's shock front, as determined
from the thickness of the filaments. The shaded area
is excluded as the filament width cannot be smaller than the diffusion
length \citep[c.f.][]{vink03a}.
\label{bfield}}
\end{figure}

However, the high spatial resolution images by \chandra\ reveal thin
filaments surrounding \casa, which are especially prominent
in images extracted in the 4-6~keV continuum dominated
band  (Fig.~\ref{bfield}). These filaments mark the onset of
radio emission \citep{gotthelf01a}, which is a clear indication that
the X-ray emission is coming from a region close to the shock front.
The spectra of these filaments
show a lack of line emission compared to neighboring
regions, an indication for a synchrotron emission component \citep{vink03a}.

The importance of these filaments is that their widths can be used to estimate
simultaneously the downstream magnetic field strength and the electron energy
\citep{vink03a,berezhko04a}. The reason is that as the plasma moves away
from the shock front with a velocity $u = \frac{1}{\chi}V_s$, 
the high energy electrons suffer synchrotron losses on a time
scale $\tau_{loss} = 635 /B^2E$~s
($V_s$\ is the shock velocity and $\chi$\ the shock compression ratio).
As a result, the highest energy electrons can only be found in a region
confined to a layer  with a thickness corresponding to the
advection length scale:
\begin{equation}
l_a = u\tau_{loss}. \label{adv}
\end{equation}
On the other hand the photon energy for synchrotron
radiation is:
\begin{equation}
\epsilon = 7.4\ E^2B {\rm \ \,keV}.\label{synch}
\end{equation}
Using the observed thickness of the filaments, 1.5\arcsec\ to 4\arcsec,
and assuming a shock compression of $\chi=4$,
one can use the above expression to show that the magnetic field strength
is $B \sim 0.1$~mG, and the maximum electron energy $\sim10-40$~TeV 
\citep{vink03a}.
This is graphically shown in Fig.~\ref{bfield}.

This result is interesting, because it establishes that the downstream magnetic
field is much higher than the compressed mean Galactic Field, 
$B_{Gal}\sim3~\mu$G, and either suggests that the magnetic field surrounding
\casa\ is high, or that indeed Cosmic Ray streaming has enhanced the magnetic
field \citep{bell01}.

In fact, it has now been established that all young supernova remnants,
\snts, Tycho's, and Kepler's SNR have thin non-thermal X-ray filaments,
indicating similarly high magnetic fields \citep{bamba03,hwang02,cassam04}. 
As it is hard to believe that
the magnetic field is high in all young Galactic SNRs,
this is evidence for cosmic ray induced magnetic field amplification.

Shortly after the publication of the measurement of \casa's magnetic
field, a different method to estimate the magnetic
field near the shock front of \snts\ was published 
\citep{yamazaki04}. 
Instead of the advection length
the diffusion length was used 
for the length scale defining the widths of the filaments.
A similar method was also used by Berezhko et al. for both \snts\ and \casa\
\citep{berezhko03c,berezhko04a}. They 
argue  that, due to projection effects,
the actual widths are even smaller than the
observed widths, and as a consequence $B\sim0.5$~mG for \casa.

Projection effects may, indeed, be important, but one should be cautious,
since for this study \casa's brightest filaments 
were selected, and it may be that the filaments
are only locally very bright, in which case a spherical shell model may not be 
appropriate. For \snts\ there is evidence
that the non-thermal emission is not axi-symmetric, but confined
to two polar caps \citep{willingale96,rothenflug04}.

Concerning whether the diffusion or the advection model is right. 
The distinction for loss limited electron spectra is actually not
relevant, they give the same results.
I have illustrated this in Fig.~\ref{bfield}, where the shaded area shows
the values of electron energy and B-field that are excluded, based
on the fact that the filament width should be equal to, or larger than 
the diffusion length, $l_d$. 
I have assumed here Bohm diffusion:
\begin{equation}
l_d = \frac{\kappa}{u} = \frac{1}{3}\frac{cE}{eB}\frac{1}{u}, \label{ldiff}
\end{equation}
with $\kappa$ the diffusion coefficient.
That is, the boundary of the
region is formed by the diffusion length scale, and this gives 
a solution almost identical to using 
the advection length scale $l_a$\ (eq.~\ref{adv}).
This is apparently not only true for \casa\ but for all young remnants
\footnote{As shown by Dr. J. Ballet at the COSPAR 2004 convention in Paris.}, 
i.e. it looks like
$l_a \approx l_d$. 
Figure~\ref{bfield} does not take into account projection effects, but still
indicates a slightly higher magnetic field of $B = 0.2$~mG and $E = 25$~TeV.

It is not difficult to see as to why  $l_a \approx l_d$\ in
the framework of first order Fermi acceleration.
First of all note that the acceleration time, $\tau_{acc}$, is approximately
\begin{equation}
\tau_{acc} = \frac{\kappa}{u^2}.\label{acc}
\end{equation}
For efficient acceleration $\tau_{acc} < \tau_{loss}$, i.e. 
(eq.~\ref{adv},\ref{acc})
\begin{equation}
\frac{\kappa}{u^2} < \frac{l_a}{u}\
\Longleftrightarrow\  \frac{\kappa}{u} = l_d < l_a.
\label{ineq}
\end{equation}

In other words the advection length scale has to be larger than the diffusion
length scale. As soon as for a given energy $l_d \approx l_a$,
efficient acceleration stops, and the electron spectrum will be 
cut off, resulting in a steepened photon spectrum
around the corresponding photon energy. 
This is in agreement with the X-ray spectrum,  as the observed spectral 
energy index
is $-2$, steeper than the radio spectral index
of $-0.78$.

This result has some interesting consequences. First of all, note
that the steep spectral index in combination with the observation that
$l_a \approx l_d$, is additional proof for
the correctness of the explanation of the X-ray synchrotron filaments. 
Secondly, we find $l_a \approx l_d$ using the
approximation of Bohm diffusion. 
Without Bohm diffusion we would find $l_d > l_a$, in violation
of (\ref{ineq}).
The observations 
therefore, support the idea that the diffusion
is near or at the Bohm diffusion limit ($\delta B/B \sim 1$)
\citep[c.f.][]{vink04b}, 
as is expected for non-linear magnetic field amplification.
And thirdly, the electron spectrum is apparently loss limited, and not age
limited. For an age limited spectrum the maximum energy for the electrons
should be similar to that of the proton energy, but for a loss limited
spectrum the nuclei can obtain higher energies.
Using the shock properties of \casa\ and $B=0.2$~mG, the maximum
energy for nuclei can be $E_{max} \sim 2\times10^{15}Z$~eV,
close to the ``knee''.
Note, however, that the magnetic field may have been higher in the past.

\begin{figure}
\hbox{

  \psfig{figure=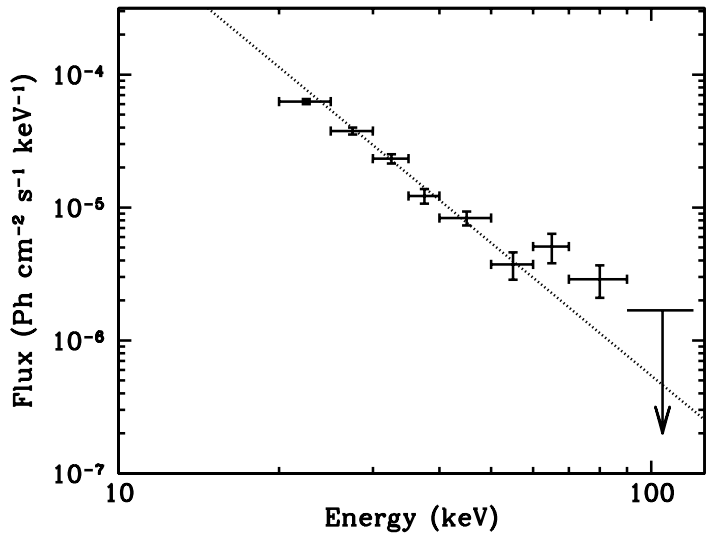,angle=-90,height=0.5\textwidth}
}
  \caption{
A hard X-ray spectrum of \casa\ as recently observed with the
\intg-IBIS instrument (Vink et al. in preparation). 
The dotted line indicates the best fit
power law spectrum as previously determined using \sax\ data \citep{vink01a}.
The excess around 75~keV is due to line emission caused by the decay of 
\tiff\ \citep{vink01a}.
\label{ibis}
}
\end{figure}

\section{The nature of \casa's hard X-ray emission}
One of the first arguments in favor of 
X-ray synchrotron radiation from \casa\
was the existence of non-thermal, hard X-ray emission 
\citep{the96,allen97,favata97} 
(Fig.~\ref{ibis}).
However, the non-thermal rims observed by \chandra\ are likely to contribute
only a fraction of 10\% to the total flux. 

Unfortunately, imaging above 15~keV is extremely difficult, and our
knowledge of where the hard X-ray emission predominantly comes from
is based on the highest energies that can be imaged with
\sax\ and \xmm, i.e. the 8-15 keV energy band \citep{vink99,bleeker01}.
These images, however, suggest that the hard X-ray emission is associated
with the shell of \casa\ and peaks in the West \citep{vink99,bleeker01}.
This is also the location of a number of continuum
dominated filaments seen by \chandra\ \citep[e.g.][]{hughes00a}.

One could of course explain this by assuming that the filaments
are similar to the filaments at the edge of \casa, but projected
onto the interior. There are, however, problems with that.
First of all, the outside filaments probably just look like filaments
due to strong limb brightening.
The other argument is that \chandra\ proper motion studies indicate that
the kinematics of the interior filaments are systematically different
from the kinematics of the outer filaments \citep{delaney04}. 
The latter have a dynamical
timescale of $\sim 500$~yr, whereas the interior filaments have a timescale
similar to the radio knots, i.e. $>800$~yr \citep{anderson95}.

So what can explain both the presence of the interior filaments,
and the hard X-ray spectrum?
One problem with X-ray synchrotron radiation is the extremely short life time
of the electrons responsible for it: assuming for the interior 
$B \sim 1$~mG, $\tau_{loss} \sim 2$~yr. This would require (re)acceleration
in the interior of the remnant. Another problem is that the spectral
slope seems to remain constant at $\Gamma = 3.3$, 
from 6-60 keV, whereas X-ray synchrotron radiation is expected to have
gradually steepen.

An explanation may be that the hard X-ray emission is due
to non-thermal bremsstrahlung from supra-thermal
electrons with energies up to 100~keV.
A possible model is electron acceleration by lower hybrid waves 
\citep{laming01a,laming01b}, which predicts the correct spectral slope
\citep{vink03a}.
However, some of the filaments seem to lack line emission \citep{hughes00a}, 
which is a problem
for non-thermal bremsstrahlung models. 

One way to reconcile the idea of X-ray synchrotron radiation with
the high interior magnetic field strength is to invoke those high magnetic
fields. It is likely that \casa's bright radio ring is associated
with the contact discontinuity between swept up matter and ejecta.
The contact discontinuity is predicted to have a high density, and
the magnetic fields may be enhanced due to Rayleigh-Taylor instabilities.
It is not unreasonable to suppose that electrons diffusing from a low
magnetic field region into the high field region near the contact
discontinuity flare up \citep{lyutikov04}. 

In fact, electrons entering the high magnetic field
will gain energy due to the betatron effect \citep[e.g.][]{kirk94}.
As an example, assume that the the plateau of \casa\ has a magnetic
field of 0.5~mG, whereas certain regions have magnetic fields
as high of 2~mG \citep{wright99}. If the electrons have been accelerated
100~yr ago, the maximum energy they can have is $\sim 5$~erg,
having a peak emission at 37~eV.
As the electrons enter the 2~mG region the betatron mechanism
can increase the energy by maximally a factor 2.
The peak radiation energy, however,  can increase
by a factor 16 (see eq.~\ref{synch}).
This does require that the electrons diffuse rapidly: 
a 10~erg electron in a 2~mG field has a half energy life of 6 months.

Diffusion of electrons into high magnetic field regions may
explain why the X-ray filaments seem to mark
the edges of bright radio regions \citep{delaney04}: 
the highest energy electrons cannot penetrate the
high magnetic field regions too deeply without losing a substantial
part of their energy. Moreover, the non-thermal emission may be more
pronounced in the West, as the bright ring appears closer to the shock front.
In other words, the electrons may have been accelerated more recently,
and as a consequence they start with a higher energy in the first
place.

\begin{figure}
\vbox{
\hbox{\centerline{
  \psfig{figure=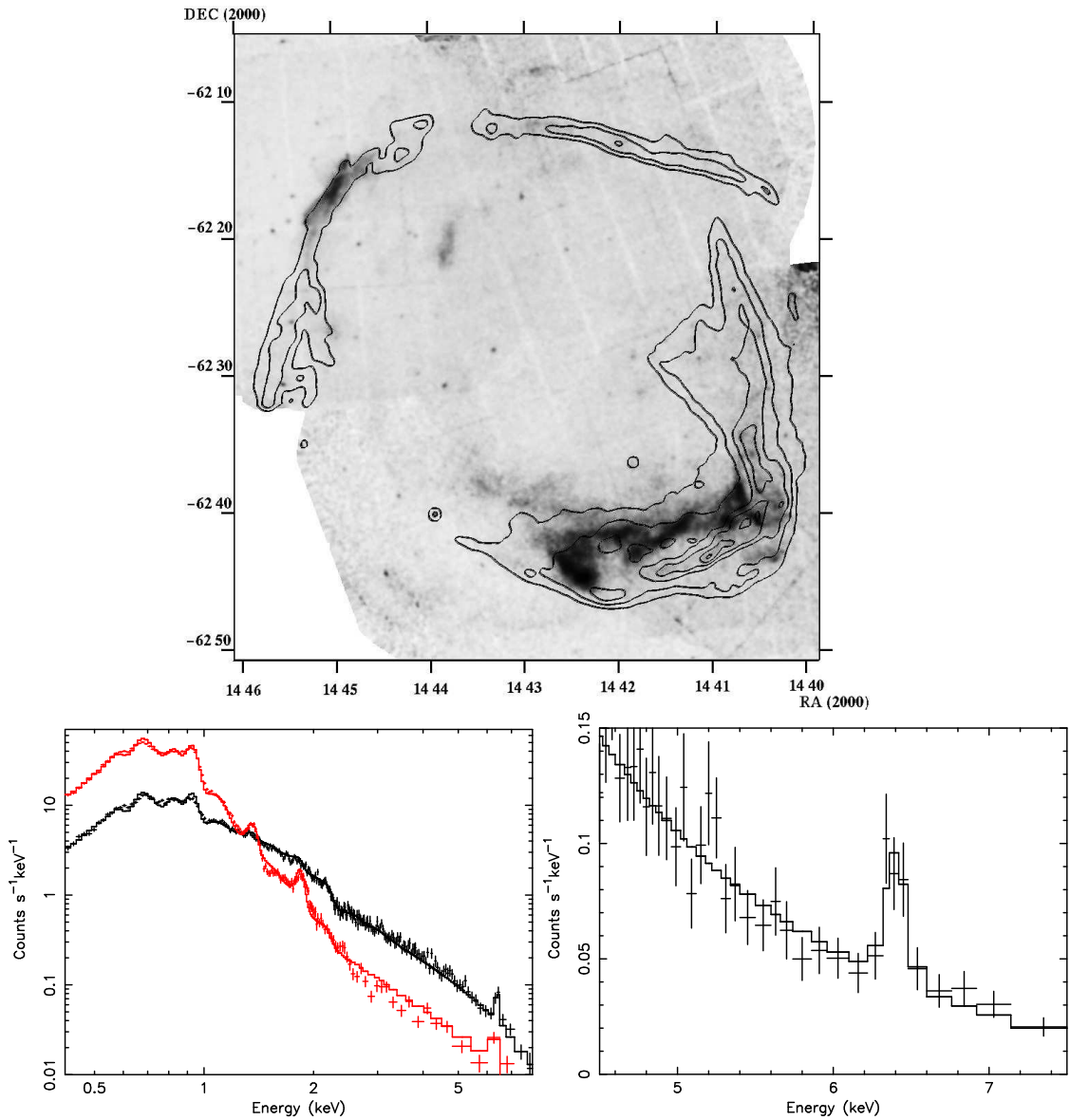,width=0.65\textwidth}
}
}
\hbox{
  \psfig{figure=rcw86_sw_hard_n_soft.ps,angle=-90,width=0.5\textwidth}
  \psfig{figure=rcw86_south_pn_hard_fek.ps,angle=-90,width=0.5\textwidth}
}}
  \caption{
Left:
\xmm\  
X-ray mosaic image of \rcwes\ in the energy band 1.95-6.8 keV with softer
0.5-1.0 keV emission 
overlayed as contours. 
Right, top: Spectra from two distinct regions, one dominated  by thermal
emission, the other by non-thermal emission.
Right, bottom: Detail of the non-thermal spectrum from the Southwest,
showing the Fe-K line at 6.4~keV.
\label{rcw86}
}
\end{figure}

\section{Diffuse X-ray synchrotron emission: the case of \rcwes}
The X-ray synchrotron emission from the narrow outer rims of \casa\ and other
young remnants can be understood in the framework of 
diffusive shock acceleration in combination with synchrotron losses.
Ironically, however,  the X-ray synchrotron emission from \rxjSNR,
which thanks to observations by
\cang\ and \hess\ may now become an icon of a cosmic
ray accelerating SNR \citep{enomoto02}, 
seems much more complicated \citep{uchiyama03}.

In many respects \rcwes\ (G315.4-2.3) is similar to \rxjSNR;
it is a large
remnant of 42\arcmin, surrounded by molecular clouds observed
in CO (Dr. Y. Moriguchi, private communication). 
Most importantly, its non-thermal X-ray emission
is relatively diffuse, 
but unlike \rxjSNR\ there is substantial thermal X-ray emission.
The morphology of the thermal X-ray emission is rather different
from the non-thermal X-ray emission \citep{vink97} (Fig.~\ref{rcw86}).
The thermal emission comes from a relatively thin, curving shell, 
whereas the hard X-ray emission comes predominantly from a slab interior
to the southwestern shell, an isolated patch in the the North, 
and from an isolated patch associated with the northeastern shell.

One puzzling feature of the non-thermal X-ray emission of \rcwes\ is 
the presence of 6.4~keV line emission \citep{vink97,bocchino00,bamba00,rho02}, 
associated with
iron in a low ionization state ($<$ Fe XVII, Fig.~\ref{rcw86}).
It was initially proposed that the Fe-K emission pointed to a non-thermal
bremsstrahlung of the continuum \cite{vink97}. In this explanation the
continuum and the line emission are caused by the same electrons with energies
in the 1-100 keV range. However, a non-thermal electron distribution cannot
be maintained for a long time. For that reason it was proposed that the
continuum emission is X-ray synchrotron radiation, whereas the Fe-K emission
comes from a hot, underionized, pure Fe plasma 
\citep{borkowski01,bamba00,rho02}. The plasma needs to be
pure Fe in order to suppress the bremsstrahlung continuum with respect to
the line emission. I will adapt here the interpretation that the non-thermal
X-ray emission is synchrotron radiation, but I like to point out that
an alternative interpretation of the Fe-K emission is 
fluorescence caused by low energy ($\sim$ MeV) protons. An elaboration on
the Fe-K is beyond the present discussion, and will be published elsewhere.

\rcwes\ seems to be the result of an explosion in an OB association at a 
distance of 2.5~kpc \citep{westerlund69,rosado94}.
It is therefore likely that the remnant develops in a cavity blown
by the hot stars of the OB association \citep{vink97,dickel01}. 
This can explain the curvy
nature of the thermal shell, which may be the results of overlapping
stellar wind bubbles.
In the Southwest the shock  seems to run into
a molecular cloud with a density of $n~\sim10$~cm$^{-3}$.

The shock velocity of \rcwes\ has been measured 
using the H$\alpha$\ line widths,
indicating $V_s\sim400-800$~\kms\ \cite{ghavamian01}.
This is is a factor 5 to 10 lower than the shock velocity
of remnants like \casa. It is therefore very surprising that
\rcwes\ emits X-ray synchrotron emission, as
the maximum photon energy for synchrotron
radiation is independent of the magnetic field and
scales with $\propto V_s^2$\ \citep{uchiyama03}.

Here I discuss preliminary results based on recent \xmm\ and \chandra\
observation of the northeastern shell of \rcwes.
The main reason to limit the present discussion to the 
northeastern part is that at least for that region one can
be certain that the forward shock is responsible for the shock acceleration
of the X-ray synchrotron emitting electrons,
whereas for the other X-ray synchrotron radiating regions
either the forward shock is responsible, but the X-ray synchrotron radiation
is projected interior to the main shell,
or the
electrons have been accelerated by the reverse shock \citep{rho02}.

\begin{figure}
\hbox{
  \vbox{
  \psfig{figure=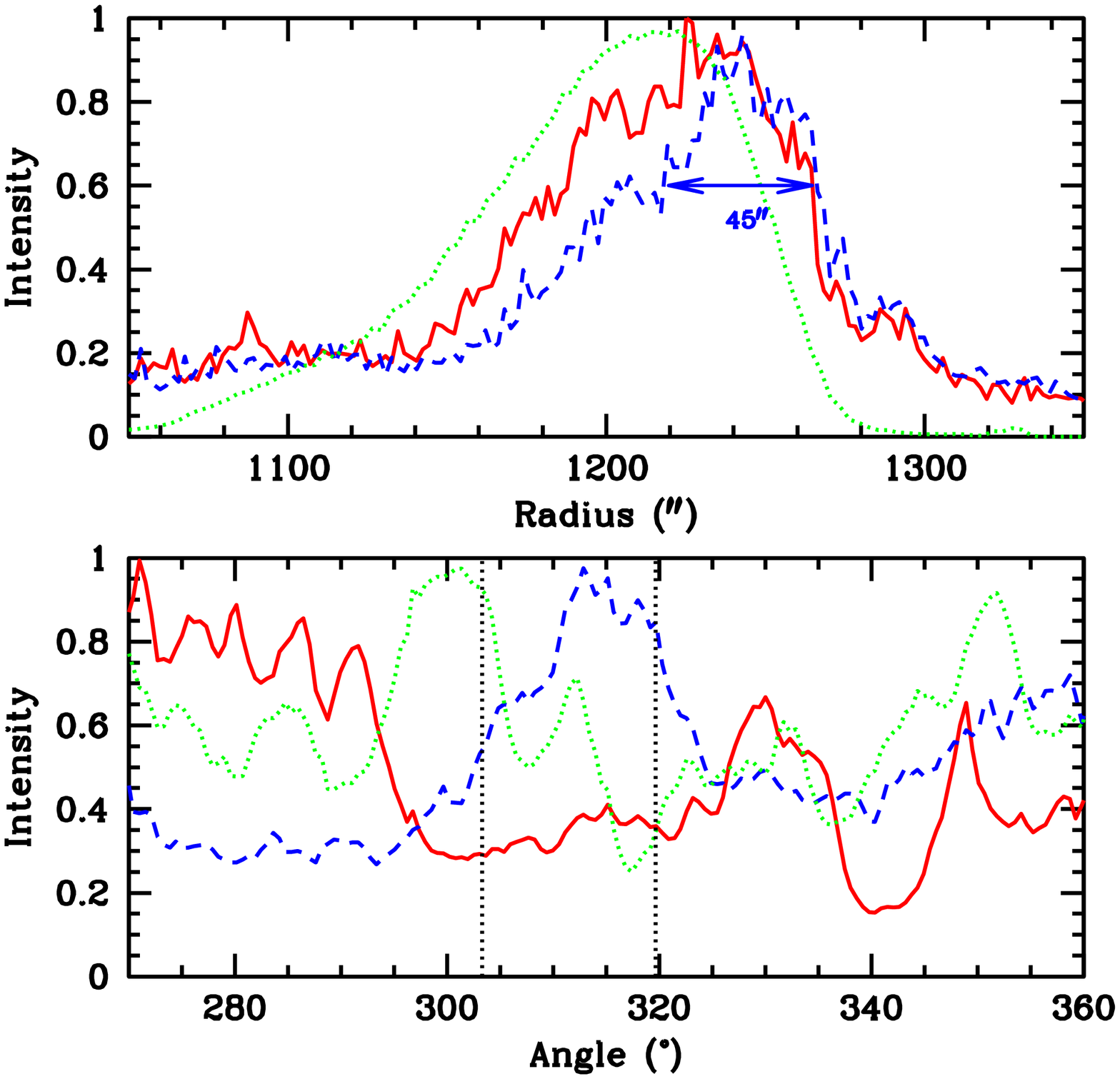,width=0.55\textwidth}
 \vskip 5mm
}
\vbox{
  \psfig{figure=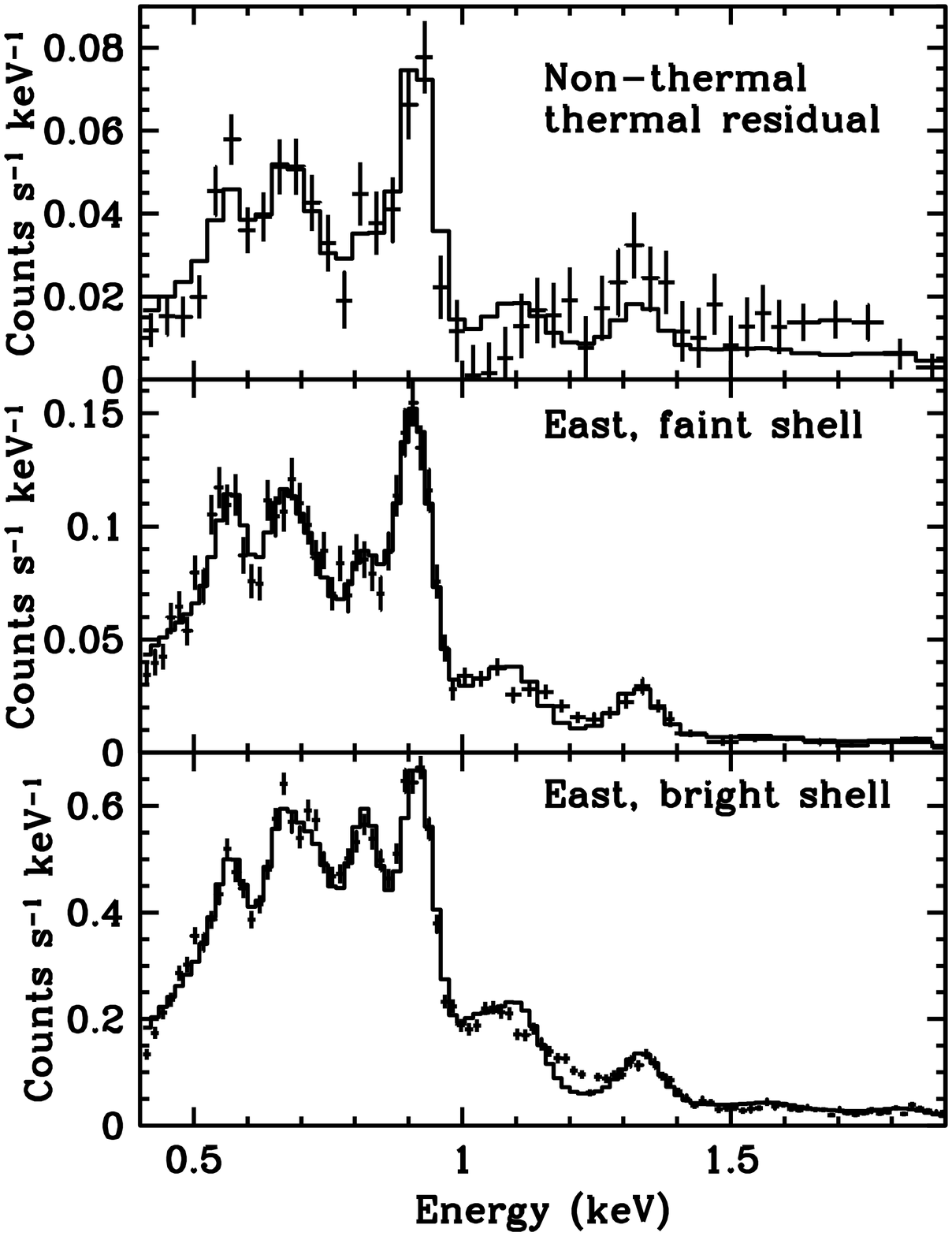,angle=0,width=0.45\textwidth}
 }
}
  \caption{
Right: Intensity profiles of the northeastern part of \rcwes, 
both in the radial (top) and
tangential direction (bottom, from East to North). 
The 0.5-1 keV X-ray emission
is indicated by a red, solid line, the 2-6.8 keV emission by
a blue, dashed line and the radio emission \citep{dickel01} by a green,
dotted line.
Left:  Thermal spectra from the eastern shell of \rcwes.
The upper panel shows the non-thermal spectrum after subtraction of the best 
fit power law spectrum.
Note the absence of Fe-L emission at $\sim 0.85$~keV 
in both the faint outside shell and in the residual thermal emission.
\label{rcw86ne}
}
\end{figure}

\section{Non-thermal X-rays from the Northeast of RCW 86}
Figure~\ref{rcw86ne} shows the emission profiles for the northeastern
shell of \rcwes, both for the radial and tangential direction.
It is clear from the radial profiles that the X-ray synchrotron emission
is more confined to the edge of the shell than the radio and thermal
X-ray emission, as is to be expected for the emission
from ultra-relativistic electrons accelerated by the forward shock.
The typical width of the synchrotron region is $\sim 45$\arcsec\
($l = 1.7\times10^{18}$~cm). Following a similar
line of reasoning as for \casa\ one can calculate the magnetic field
using either the advection (eq.~\ref{adv}) or the diffusion length
scale (eq.~\ref{ldiff}).
However, these produce inconsistent answers.
For instance, allowing for some uncertainty due to projection
effects, one finds for $V_s = 600$~\kms\ 
$B = 45-90~\mu$G using the diffusion length scale, but this implies
a much smaller advection scale of $l_a = 9\times10^{16}$~cm,
violating the requirement for efficient acceleration $l_d < l_a$ (\ref{ineq}).
Similarly, using the advection length scale gives $B = 6-13~\mu$G,
which implies a diffusion length of $l_d = (1-3)\times10^{19}$~cm,
again in violation of (\ref{ineq}).

The emission profiles along the shell are also surprising,
as the thermal and non-thermal X-ray emission are anti-correlated;
the non-thermal emission seems to fill in a gap in the soft X-ray shell.
One may think that this points toward locally very efficient cosmic ray
acceleration at the cost of heating the plasma. However, one would
then also expect the radio emission to be bright, wherever
the non-thermal X-ray emission is bright, but this is clearly not the case
(Fig.~\ref{rcw86ne}).

Moreover, the thermal emission from this region is very similar to
that of the rest of the shell. The thermal emission along the shell
shows variation, which is most likely predominantly caused by variations
in the ionization parameter 
\net.\footnote{The thermal emission from SNRs is often out
of ionization equilibrium.
The ionization process scales with
the product of electron density and time.} 
The residual thermal
emission from the hard X-ray emission is similar to the spectra 
of other faint regions that have low ionization parameters,
i.e. \net\ $= 4\times10^{9}$~\netunit\ for $kT \sim 1$~keV.
This suggests that the non-thermal X-ray emission is coming from a region
with a low density. 

\begin{figure}
\psfig{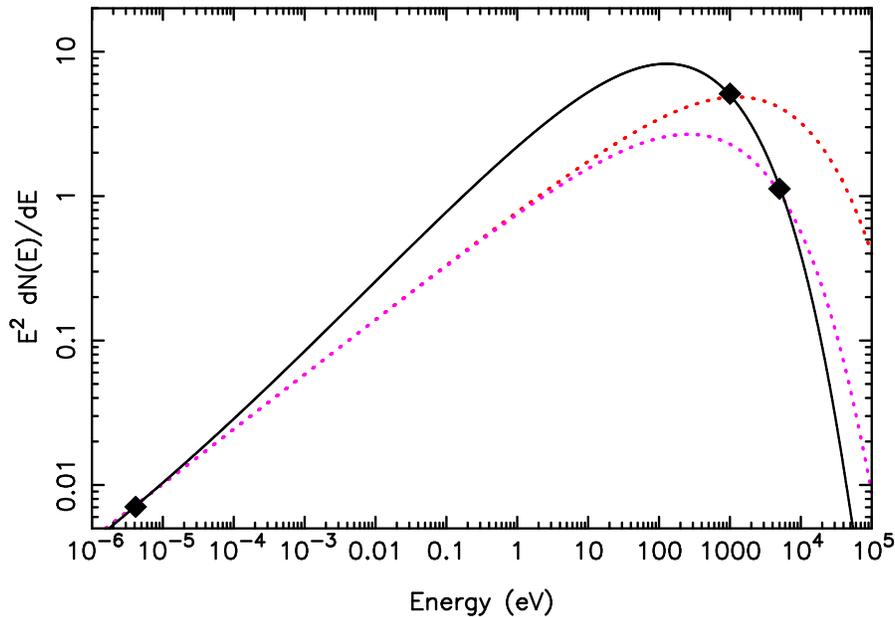}
\caption{
Synchrotron model for the hard X-ray emitting region in the Northeast.
The solid line assumes an electron spectrum that flattens with energy,
the colored lines assume a constant power law index (particle index -2.2).
\label{rcw86model}
}
\end{figure}

Interestingly, the radio emission from the region with the non-thermal
X-ray emission is rather weak, and can only be jointly fitted with
a synchrotron model if a flattening of the electron spectrum 
from -2.2 to 2 is assumed 
(Fig.~\ref{rcw86model}). This is in agreement with theoretical
calculations of cosmic ray spectra 
that take into account the effect of the cosmic ray pressure on the shock
structure \citep[e.g.][]{ellison91}.

So how may the observational facts help to explain the emergence
of non-thermal X-ray emission at this particular spot?
A solution to this problem may be the 
large density gradients
that are present in SNRs evolving
in a cavity blown by stellar winds. 
Part of the shock may already be plowing through
the dense cavity wall, whereas in other regions
the shocks still move through the tenuous cavity itself,
producing only weak, thermal X-ray emission.
As a result, locally, the shock velocity may be substantially
higher than $V_s= 600$~\kms, as the H$\alpha$ measurements
are biased toward bright, dense, filaments. 
Note that for cavity remnants
the shock velocity remains high while the shocks move through the cavity, 
and
decelerates rapidly once interacting with the shell \citep{tenorio91}.

We can estimate the shock velocity and magnetic field strength by
demanding that the advection scale and diffusion length scale
are equal and consistent with the observed width of the non-thermal
X-ray emission of $\sim45$\arcsec.
This is the case for $V_s \approx  2700$~\kms, 
which would imply $B \approx 17~\mu$G. This is an order of magnitude lower
than for \casa\ or SN~1006 and helps to explain why the non-thermal
X-ray emission is rather diffuse.

Next consider whether such a variation in shock velocity
is consistent with the morphology of the  northeastern shell, i.e.
is the lack of pronounced bulging out of the shell consistent with
a locally higher shock velocity?
To answer that we need to estimate the time since the brighter parts
of the eastern shell started interacting with the cavity wall. 
For an order of magnitude estimate
we can use the diagnostics offered by the thermal spectra,
which indicate \net $=5.4\times10^9$~\netunit, whereas the emission
measure together with a volume estimate indicates 
$n_{\rm e} \sim 0.5$~cm$^{-3}$.
Together they suggest that the bulk of the plasma was shocked
$\Delta t = 340$~yr ago. 
Therefore the shock that produces the X-ray synchrotron emission
could be as far ahead as $\Delta r = \Delta V_s  \Delta t =  0.7$~pc,
corresponding to a fractional difference of $\Delta r/r = 0.7/16 = 0.04$,
which is very reasonable given \rcwes's morphology.

The idea of substantial local variations in the shock
velocity can be best tested by measuring the shock velocity, either
from the faint H$\alpha$ filaments near the X-ray synchrotron region, or
even better, by directly measuring the proper motion of the edge of the
non-thermal X-ray emitting region.

\section{Summary}
The discovery of X-ray synchrotron radiation from shell-type supernova 
remnants has not only provided evidence that supernova remnant shocks
are efficient particle accelerators, but has given us a new tool
to study the acceleration process.
In that respect the high spatial resolution of \chandra\ has been crucial,
as it shows that all young Galactic 
remnants have narrow X-ray synchrotron emitting filaments.
The spectra from, and the widths of these filaments indicate that:

\begin{itemize}
\itemsep 0mm
\item the electron cosmic ray spectrum is loss limited,
\item the downstream magnetic fields are relatively high, $B \geq 0.1$~mG, 
and consistent with
magnetic field amplification by cosmic ray streaming,
\item the maximum electron energy is $10-40$~TeV,
\item diffusion takes place at, or close to, the Bohm limit.
\end{itemize}
The maximum electron energy falls short of the ``knee'' energy.
However, protons are not limited by synchrotron losses. Moreover,
the Bohm diffusion and the high magnetic fields makes a maximum proton energy
as high as the ``knee'' plausible, certainly if in the very early phases
of these remnants the shock velocities and magnetic fields were higher.

Although the X-ray synchrotron radiation from the narrow filaments
of young remnants can be well understood within the framework of
diffusive shock acceleration, the more extended X-ray synchrotron radiation
from large remnants such as  \rxjSNR\ and \rcwes\ poses some difficulties.
I have discussed the X-ray synchrotron emission from a shock region
in the Northeast of \rcwes, and I have argued that it can be best understood
if the magnetic field is $B \approx 17~\mu$G, and the X-ray synchrotron
radiation is comes from region where the shock velocity
is considerably faster ($V_s \approx 2700$~\kms) 
than elsewhere along the shell ($V_s \approx 600$~\kms).

\begin{theacknowledgments}
I would like to thank my collaborators 
Johan Bleeker, Martin Laming, Peter den Hartog, and Andrei Bykov
for their contributions and help, and 
John Kirk for discussing with me betatron acceleration as a 
mechanism to enhance X-ray synchrotron emission in \casa.
\end{theacknowledgments}

\end{document}